\documentclass[aps,prd,superscriptaddress,showpacs,preprint]{revtex4}
\usepackage{graphicx, bm}
\usepackage[usenames]{color}

\begin{document}
\draft
\title{Bounds on the electromagnetic dipole moments through the single top production at the CLIC}

\author{M. K\"{o}ksal\footnote{mkoksal@cumhuriyet.edu.tr}}
\affiliation{\small Deparment of Optical Engineering, Cumhuriyet University, 58140, Sivas, Turkey.\\}

\author{A. A. Billur\footnote{abillur@cumhuriyet.edu.tr}}
\affiliation{\small Deparment of Physics, Cumhuriyet University, 58140, Sivas, Turkey.\\}

\author{ A. Guti\'errez-Rodr\'{\i}guez\footnote{alexgu@fisica.uaz.edu.mx}}
\affiliation{\small Facultad de F\'{\i}sica, Universidad Aut\'onoma de Zacatecas\\
         Apartado Postal C-580, 98060 Zacatecas, M\'exico.\\}

\date{\today}

\begin{abstract}

We obtain bounds on the anomalous magnetic and electric dipole moments of the $t$-quark from a future
high-energy and high-luminosity linear electron-positron collider, as the CLIC,
with polarized and unpolarized electron beams which are powerful tools for determining new physics.
We consider the processes $\gamma e^- \to \bar t b\nu_e$ ($\gamma$ is the Compton backscattering photon)
and $e^+e^- \to e^-\gamma^* e^+ \to \bar t b\nu_e e^+$ ($\gamma^*$ is the Weizsacker-Williams photon)
as they are one of the most important sources of single top-quark production. For systematic uncertainties
of $\delta_{sys}=0\%\hspace{1mm}(5\%)$, $b-\mbox{tagging efficiency}=0.8$, center-of-mass energy of
$\sqrt{s}=3\hspace{0.8mm}TeV$ and integrated luminosity of ${\cal L}=2\hspace{0.5mm}ab^{-1}$ the future
$e^+e^-$ collider may put bounds on the electromagnetic dipole moments $\hat a_V$ and $\hat a_A$ of the top
quark of the order of ${\cal O}(10^{-2}-10^{-1})$ at the $2\sigma\hspace{1mm}(3\sigma)$ level, which are
competitive with those recently reported in previous studies at hadron colliders and the ILC.
\end{abstract}


\vspace{5mm}

\maketitle

\section{Introduction}

The top quark is by far the heaviest particle of the Standard Model (SM) \cite{Glashow,Weinberg,Salam},
with a  mass of $m_t=173.5\pm 0.6\hspace{1mm} (\mbox{stat.}) \pm 0.8\hspace{1mm} (\mbox{syst.})$ \cite{Data2014}.
Up to now, the top quark has only been studied at the Tevatron and Large Hadron Collider (LHC). Its large
mass implies that the top quark is the SM particle most strongly coupled to the mechanism of electroweak symmetry
breaking. This is the principal reason it is considered to be one of the most likely places where new physics might
be discovered. This means the top quark is a window to any new physics at the $TeV$ energy scale. While much information
about the top quark is already available that shows consistency with SM expectations, its properties and interactions are
among the most important measurements for present and future high energy colliders \cite{ATLAS,CMS,Abe,Aarons,Brau,Baer,Accomando,Dannheim,Aguilar0}.

The construction of a high-energy $e^+e^-$ International Linear Collider (ILC) has been proposed to complement direct
searches carried out at the LHC. Precision measurements of top quark properties, in particular
of its couplings, are especially interesting because the top quark is the heaviest known elementary particle and thus expected to
be more sensitive to new physics at higher scales.

The top quark has been studied in some detail at the Tevatron and LHC. Many of its properties are still poorly constrained such as mass, spin, color and electric charges, the electric and magnetic dipole moments and the chromomagnetic and chromoelectric dipole
moments. Therefore, significant new insights on top quark properties will be one of the tasks
of the LHC, the ILC \cite{Abe,Aarons,Brau} and the Compact Linear Collider (CLIC) \cite{Accomando,CLIC}.

The dipole moments of the top quark are some of the most sensitive observable, and although these intrinsic
properties have been studied extensively both theoretically and experimentally, it is necessary to have
more precise measurements. The dipole moments of the top quark have been
investigated by several authors  and in a variety of theoretical models \cite{Soni,Soni1,Bartl,Hollik,Huang,Sprinberg,Roberto,Perez}.
Further, a number of studies show that in the processes $e^+e^- \to t\bar t$ and $\gamma\gamma \to t \bar t$,
the dipole moments of the top quark can be measured with great sensitivity \cite{Atwood,Polouse,Choi,Polouse1}.
However, there are a significant number of top quarks that are produced in single form via the
weak interaction. There are several single top quark production processes of interest
in $e^+e^-, e^-e^-, \gamma e^-$ and $\gamma\gamma$ collisions, characterized
by the virtuality of the W boson \cite{Boos,Batra,Fuster,Boos1,Boos2,Boos3,Jikia,Boos4,Boos5,Jun}.

Although studying single top quark production may not be considered of great importance,
there are several reasons why its study is necessary in future linear $e^+e^-$ colliders:
1) It is a very good alternative to study the dipole moments $\hat a_V$ and $\hat a_A$
of the top quark, as well as the anomalous coupling $tbW$. 2) Single top production at CLIC in association with
a $W$ boson and bottom quark through $WW^*$ production leads to the same final state as $t$
quark pair production. 3) The cross section for single top quark production processes is significant
since production is abundant in $e^+e^-$ colliders that operate at high energies.
In addition, the single top quark production is directly proportional to the square
of the $tbW$ coupling, and therefore it is potentially very sensitive to the $tbW$ structure \cite{Weiglein}.
4) Single top quarks are produced with nearly $100\%$ polarization due to the weak interaction \cite{Heinson,Carlson}.
5) New physics can influence single top production by inducing weak interactions beyond the SM weak interactions \cite{Carlson,Carlson1},
through loop effects \cite{Simmons,Atwood1,Li}, or by providing new sources of single top quark production \cite{Malkawi,Datta,Oakes}.
For these reasons, it is important to study the properties of the top quark, in particular their dipole
moments through the single top quark production processes.

In the SM, the prediction for the Magnetic Dipole Moment (MDM) of the top quark is $a^{SM}_t=0.02$ \cite{Benreuther},
which can be tested in current and future colliders, such as LHC and CLIC. In contrast, its Electric Dipole Moment (EDM)
is strongly suppressed and less than $10^{-30}\hspace{0.8mm}e{\mbox{cm}}$ \cite{Hoogeveen,Pospelov,Soni},
which is much too small to be observed. It is, however, highly attractive for probing new physics.

The sensitivity to the EDM has been studied in models with vector-like multiplets which predicted the top quark
EDM close to $1.75\times 10^{-3}$ \cite{Ibrahim}.

There are studies performed via the $t\bar t\gamma$ production for the LHC at $\sqrt{s}=14\hspace{0.8mm}TeV$ and
${\cal L}=300\hspace{0.8mm}fb^{-1}$ and $3000\hspace{0.8mm}fb^{-1}$, with limits of $\pm 0.2$ and $\pm 0.1$,
respectively \cite{Baur}. Other limits are reported in the literature: $-2.0\leq \hat a_V\leq 0.3$ and $-0.5\leq \hat a_A\leq 1.5$
which are obtained from the branching ratio and the CP asymmetry from radiative $b \to s\gamma$ transitions \cite{Bouzas},
while the bounds of $|\hat a_V|< 0.05\hspace{0.8mm}(0.09)$ and $|\hat a_A|< 0.20\hspace{0.8mm}(0.28)$ come
from measurements  of $\gamma p\to t\bar t$ cross section with $10\%$ $(18\%)$ uncertainty, respectively \cite{Bouzas1}.
More recent limits on the top quark magnetic and electric dipole moments through the process $pp \to p\gamma^*\gamma^*p \to p t\bar tp$
at the LHC with $\sqrt{s}=14\hspace{0.8mm}TeV$, ${\cal L}=3000\hspace{0.8mm}fb^{-1}$ and $68\%$ C.L. are $-0.6389\leq \hat a_V\leq 0.0233$
and $|\hat a_A|\leq 0.1158$ \cite{Sh}. Sensitivity limits for the anomalous couplings of the top quark through the production process
of top quark pairs $e^+e^- \to t\bar t$ for the ILC at $\sqrt{s}=500\hspace{0.8mm}GeV$, ${\cal L}=200\hspace{0.8mm}fb^{-1}$, ${\cal L}=300\hspace{0.8mm}fb^{-1}$ and ${\cal L}=500\hspace{0.8mm}fb^{-1}$ are predicted to be of the order of ${\cal O}(10^{-3})$.
Thus, the measurements at an electron positron collider lead to a significant improvement in comparison with LHC.
Detailed discussions on the dipole moments of the top quark in top quark pairs production at the ILC are reported in the literature \cite{Atwood,Polouse,Choi,Polouse1,Aguilar0,Amjad,Juste,Asner,Abe,Aarons,Brau,Baer}.
It is worth mentioning that there are no limits reported in the literature on the dipole moments $\hat a_V$ and $\hat a_A$ via single
top quark production processes.

CP violation was first observed in a small fraction of $K$ mesons decaying to two pions in the SM. This phenomenology
in the SM can be easily introduced by the Cabibbo-Kobayashi-Maskawa mechanism in the quark sector. For this reason, the
presence of new physics beyond the SM  can be investigated by examining the electromagnetic properties of the top quark that
are defined with CP-symmetric and CP-asymmetric anomalous form factors. Its dipole moments such as the MDM come from one-loop
level perturbations and the corresponding EDM, which is described as a source of CP violation.

Following references \cite{Sh,Kamenik,Baur,Aguilar,Aguilar1}, the definition of the general effective coupling $t\bar t\gamma$,
including the SM coupling and contributions from  dimension-six effective operators, can be parameterized
by the following effective Lagrangian:

\begin{equation}
{\cal L}_{\gamma t\bar t}=-g_eQ_t\bar t \Gamma^\mu_{\gamma t\bar t} t A_\mu,
\end{equation}

\noindent where $g_e$ is the electromagnetic coupling constant, $Q_t$ is the top quark electric charge and $\Gamma^\mu_{\gamma t\bar t}$
the Lorentz-invariant vertex function which describes the interaction of a $\gamma$ photon with two top quarks and can be parameterized by

\begin{equation}
\Gamma^\mu_{\gamma t\bar t}= \gamma^\mu + \frac{i}{2m_t}(\hat a_V + i\hat a_A\gamma_5)\sigma^{\mu\nu}q_\nu,
\end{equation}

\noindent where $m_t$ is the mass of the top quark, $q$ is the momentum transfer to the photon and the couplings
$\hat a_V$ and $\hat a_A$ are real and related to the anomalous magnetic moment and the electric dipole moment
of the top quark, respectively.

The majority of physics research in linear colliders is done assuming positron and electron beams are unpolarized.
However, another significant advantage of the linear colliders is to obtain suitability of a highly polarized electron beam
that can be polarized up to $\pm\hspace{0.8mm}80\hspace{0.8mm}\%$. A polarized electron beam provides a method to investigate the
SM and to diagnose new physics beyond the SM. Observation of even the tiniest signal which conflicts with the SM expectations
would be persuasive evidence for new  physics. Proper selection of the electron beam polarization may therefore be used to
enhance the new physics signal and also to considerably suppress backgrounds.

In this work we study the sensibility of the anomalous magnetic and electric dipole moments of the top quark through
the processes $\gamma e^- \to \bar t b\nu_e$ ($\gamma$ is the Compton backscattering photon) and
$e^+e^- \to e^-\gamma^* e^+ \to \bar t b\nu_e e^+$ ($\gamma^*$ is the Weizsacker-Williams photon) which are among
the most important sources of single top quark production \cite{Boos,Boos1}. We use center-of-mass energies of the
CLIC \cite{CLIC}. These values are for a center-of-mass energy of $1.4\hspace{1mm}TeV$  with integrated
luminosity of $1500\hspace{0.8mm}fb^{-1}$ and $3\hspace{1mm}TeV$  with ${\cal L}=2000\hspace{0.8mm}fb^{-1}$, and
polarized and unpolarized electron beams $P_{e^-}=-80\%$ and $P_{e^+}=0\%$ \cite{Moortgat}. Not only
can the future $e^{+}e^{-}$ linear collider be designed to operate in $e^{+}e^{-}$ collision mode, but it can also be
operated as a $e \gamma$ and $\gamma \gamma$ collider. This is achieved by using Compton backscattered photons in the
scattering of intense laser photons on the initial $e^{+}e^{-}$ beams. Another well-known application of linear
colliders is to study new physics beyond the SM through $e \gamma^{*}$ and $\gamma^{*}\gamma^{*}$ collisions. A quasireal
$\gamma^{*}$ photon emitted from one of the incoming $e^{-}$ or $e^{+}$ beams interacts with the other lepton shortly
after, generating the subprocess $\gamma^{*} e^{-} \rightarrow \bar t b \nu_{e}$. Hence, first we calculate the
main reaction $e^{+}e^{-} \rightarrow e^{-}\gamma^* e^{+} \rightarrow \bar t b \nu_{e}e^+$ by integrating the cross section
for the subprocess $\gamma^{*} e^{-} \rightarrow \bar t b \nu_{e}$. In this case, the quasireal photons in $\gamma^{*}e^-$
collisions can be examined by Equivalent Photon Approximation (EPA) \cite{Budnev,Baur1,Piotrzkowski}
using the Weizsacker-Williams approximation (WWA). In EPA, photons emitted from incoming leptons which have very low virtuality
are scattered at very small angles from the beam pipe. These emitted quasireal photons have a low $Q^{2}$ virtuality and are
therefore almost real. We only use the photon virtuality of $Q^2_{max} = 2\hspace{0.8mm}GeV^2$. Also, we can add parts related
to the large values of $Q^2_{max}$ which do not significantly contribute to obtaining sensitivity limits on the anomalous
couplings \cite{L3,OPAL,Sahin,Billur}. These processes have been observed phenomenologically and experimentally at the LEP,
Tevatron and LHC \cite{Abulencia,Aaltonen1,Aaltonen2,Chatrchyan1,Chatrchyan2,Abazov,Chatrchyan3,Inan,Inan1,Inan2,Sahin1,
Atag,Sahin2,Sahin4,Senol,Senol1,Fichet,Sun,Sun1,Sun2,Senol2,Atag1}.

Taking all of the aforementioned into account, we study the potential of the processes $\gamma e^- \to \bar t b\nu_e$ and
$e^{+}e^{-} \rightarrow e^{+}\gamma^{*} e^{-} \rightarrow \bar t b \nu_{e}e^+$ via Compton backscattering and WWA, respectively,
and derive bounds on the dipole moments $\hat a_V$ and $\hat a_A$ at $2\sigma$ and $3\sigma$ level ($90\%$ and $95\%$ C.L.),
and at a future high-energy and high-luminosity linear electron positron collider, as the CLIC, to study
the sensibility on the anomalous magnetic and electric dipole moments of the top quark. The corresponding schematic and Feynman
diagrams for the main reactions as well as for the subprocesses which give the most significant contribution to the total cross
section are shown in Figs. 1-2.

This paper is organized as follows. In Section II, we study the dipole moments of the top quark through the process
$\gamma e^- \to \bar t b \nu_e$ and in Section III, through the process $e^+e^- \to e^+\gamma^* e^- \to \bar t b \nu_e e^+$.
Finally, we summarize our conclusions in Section IV.

\section{Compton backscattering: Cross section of $\gamma e^- \to \bar t b \nu_e$}

In this section we present numerical results of the cross section for the process
$\gamma e^- \to \bar t b \nu_e$, using the CalcHEP \cite{Belyaev} packages for calculations of
the matrix elements and cross sections. These packages provide automatic computation of the cross sections and
distributions in the SM as well as their extensions at tree level. We consider the high-energy stage of possible
future linear $\gamma e^-$ collisions with $\sqrt{s}=1.4$ and 3$\hspace{1mm}TeV$ and design luminosity 50, 300, 500,
1000, 1500 and 2000$\hspace{1mm}fb^{-1}$ according to the new data reported by the CLIC \cite{CLIC}. In addition, in
all numerical analysis we consider the $b$-tagging efficiency of $0.8$, systematic uncertainty of $\delta_{sys}=0\%, 5\%$
and the acceptance cuts will be imposed as $|\eta^ b|< 2.5$ for pseudorapidity, $p^b_T > 20$ $GeV$ and $p^{\nu_e}_T > 10$ $GeV$
for transverse momentums of the final state particles. We also consider the hadronic decay channels of the top quark
$BR=0.676$ (Hadronic branching ratio). There are systematic uncertainties for hadron colliders for single top quark production \cite{ATLAS1}.
For example, these uncertainties arise from luminosity, jet identification, backgrounds,
$b-\mbox{tagging efficiency}$, etc.. On the other hand, linear colliders have less uncertainties with respect to
hadron colliders for determination of the cross section of single top quark production \cite{OPAL1}. Therefore, for
events estimation in $\chi^{2}$ analysis, we have taken into account $b-\mbox{tagging efficiency}$ as well as consider
systematic uncertainties of $0\%$ and  $5\%$. The values close to this systematic uncertainty value have been taken
into account in previous studies, for example in Ref. \cite{Martinez}, a $3\%$ systematic error in the
total cross section has been assumed for the $e^{-}e^{+} \rightarrow t \bar{t}$ process at the ILC. It can seen that the systematic
error in the cross section determination has been lowered from $3\%$ to $1\%$ \cite{Martinez1}. However, since there
is no study related to the systematic error on the single top quark production at the CLIC, we use systematic errors
of $0\%$ and $5\%$ for the processes studied in this paper.

In our study we examined the projected $2\sigma$ and $3\sigma$ sensitivities on the dipole moments $\hat a_V$
and $\hat a_A$ of the top quark for the processes $\gamma e^- \to \bar t b\nu_e$ ($\gamma$ is the Compton
backscattering photon) and $e^+e^- \to e^-\gamma^* e^+ \to \bar t b\nu_e e^+$ ($\gamma^*$ is the Weizsacker-Williams
photon) at the CLIC-1.4 $TeV$ and CLIC-3 $TeV$, respectively. We use the chi-squared distribution test defined as

\begin{equation}
\chi^2=\biggl(\frac{\sigma_{SM}-\sigma_{NP}(\hat a_V, \hat a_A)}{\sigma_{SM}\delta}\biggr)^2,
\end{equation}

\noindent where $\sigma_{NP}(\hat a_V, \hat a_A)$ is the total cross section including contributions from the SM
and new physics, $\delta=\sqrt{(\delta_{st})^2+(\delta_{sys})^2}$, $\delta_{st}=\frac{1}{\sqrt{N_{SM}}}$
is the statistical error, $\delta_{sys}$ is the systematic error and $N_{SM}$ is the number of signal expected
events $N_{SM}={\cal L}_{int}\times BR \times \sigma_{SM}\times \epsilon_b$ where $\epsilon_b=0.8$ is the $b-\mbox{tagging efficiency}$
and ${\cal L}_{int}$ is the integrated CLIC luminosity.

\subsection{Top quark dipole moments through the process $\gamma e^- \to \bar t b \nu_e$
with polarized and unpolarized beams}

With polarized beams of electrons and positrons, the cross section of a process can be expressed as \cite{Moortgat}

\begin{eqnarray}
\sigma(P_{e^-},P_{e^+})=&&\frac{1}{4}[(1+P_{e^-})(1+P_{e^+})\sigma_{++}+(1-P_{e^-})(1-P_{e^+})\sigma_{--}\nonumber\\
&&+(1+P_{e^-})(1-P_{e^+})\sigma_{+-}+(1-P_{e^-})(1+P_{e^+})\sigma_{-+}],
\end{eqnarray}

\noindent where $P_{e^-} (P_{e^+})$ is the polarization degree of the electron (positron) beam, while $\sigma_{-+}$
stands for the cross section for completely left-handed polarized $e^-$ beam $P_{e^-}=-1$ and completely right-handed
polarized $e^+$ beam $P_{e^+}=1$, and other cross sections $\sigma_{--}$, $\sigma_{++}$ and $\sigma_{+-}$ are defined
analogously.

The corresponding Feynman diagrams for the process $\gamma e^- \to \bar t b \nu_e$ that give the most important contribution
to the total cross sections are shown in Fig. 2. In this figure the Feynman diagrams (1)-(3) correspond to the contribution of
the SM, while diagram (4) corresponds to the anomalous contribution, i.e., for the $\gamma e^-$ collisions there is SM background
at the tree level so the total cross section is proportional to
$\sigma_{Tot}=\sigma_{SM}+\sigma_{Int}(\hat a_V, \hat a_A)+\sigma_{Anom}(\hat a^2_V, \hat a^2_A, \hat a_V \hat a_A)$,
respectively.

To illustrate our results, we show the dependence of the cross section on the anomalous couplings $\hat a_V$
and $\hat a_A$ for $\gamma e^- \to \bar t b \nu_e$ in Fig. 3 for $P_{e^-}=-80\%$, $P_{e^+}=0\%$, as well as
on unpolarized beams and two different center-of-mass energies $\sqrt{s}=1.4, 3$\hspace{0.8mm}$TeV$ \cite{CLIC},
whereas the $\hat a_V$ ($\hat a_A$) anomalous coupling is kept fixed at zero. We observed that the cross
section is sensitive to the value of the center-of-mass energies. The sensitivity to $\bar t b \nu_e$
increases with the collider energy reaching a maximum at the end of the range considered, $\hat a_{V, A}=\pm 1$,
and the cross section for $\sqrt{s}= 3\hspace{0.8mm}TeV$ increases relative to $\sqrt{s}= 1.4\hspace{0.8mm}TeV$
up to $24.5\%$ with polarized beams and up to $26.6\%$ with unpolarized beams. By contrast, in the vicinity
of $\hat a_{V, A}=0$ the total cross section is smaller. We notice that, as shown in Fig. 3, the $\gamma e^- \to \bar t b \nu_e$
production process at an CLIC-based $\gamma e^-$ collider reaches a value of $\sigma=0.55 \hspace{0.8mm}pb \hspace{2mm}(0.3\hspace{0.8mm}pb)$
for $\sqrt{s}= 3\hspace{0.8mm}TeV$ for polarized and unpolarized beams. Although the cross section for unpolarized beams
is approximately half of that of polarized beams, in both cases the $t\bar t\gamma$ coupling could be probed with remarkable
sensitivity (see Tables I, II).

\begin{table}[!ht]
\caption{Bounds on the $\hat a_V$ magnetic moment and $\hat a_A$ electric dipole moment for the process $\gamma e^- \to \bar t b \nu_e$
($\gamma$ is the Compton backscattering photon) for $P_{e^-, e^+}=-80\%, 0\%$, $b-\mbox{tagging efficiency}=0.8$, $\delta_{sys}=0\%, \hspace{2mm}5\%$
at $2\sigma$ and $3\sigma$ C.L.}
\begin{center}
\begin{tabular}{cccccc}
\hline\hline
\multicolumn{6}{c}{ $2\sigma$ C.L. \hspace{3cm}  $\delta_{sys}=0\%$  \hspace{4cm}  $\delta_{sys}=5\%$ }\\
\hline
\cline{1-6} $\sqrt{s}\hspace{0.8mm}(TeV)$  & \hspace{0.8mm} ${\cal L}\hspace{0.8mm}(fb^{-1})$  & \hspace{1cm} $\hat a_V$   & $\hspace{1cm} |\hat a_A|$
& \hspace{1cm} $\hat a_V$   & $\hspace{1cm} |\hat a_A|$ \\
\hline
1.4               &\hspace{0.8mm}  50   &\hspace{0.8mm}  [-0.1091, 0.1565]    & \hspace{1cm}   0.1615  &  \hspace{1cm} [-0.1902, 0.2406] & \hspace{1cm} 0.2352 \\
1.4               &\hspace{0.8mm}  300   &\hspace{0.8mm} [-0.0630, 0.1103]    & \hspace{1cm}   0.1032  &  \hspace{1cm} [-0.1785, 0.2289] & \hspace{1cm} 0.2235 \\
1.4               &\hspace{0.8mm}  500  &\hspace{0.8mm}  [-0.0534, 0.1007]    & \hspace{1cm}   0.0908  &  \hspace{1cm} [-0.1775, 0.2279] & \hspace{1cm} 0.2224 \\
1.4               &\hspace{0.8mm}  1000  &\hspace{0.8mm} [-0.0424, 0.0897]    & \hspace{1cm}   0.0763  &  \hspace{1cm} [-0.1767, 0.2271] & \hspace{1cm} 0.2217 \\
1.4               &\hspace{0.8mm}  1500  &\hspace{0.8mm} [-0.0369, 0.0842]    & \hspace{1cm}   0.0689  &  \hspace{1cm} [-0.1765, 0.2268] & \hspace{1cm} 0.2214 \\
\hline
3                 &\hspace{0.8mm}  50   &\hspace{0.8mm}  [-0.0724, 0.0816]    & \hspace{1cm}   0.0768  &  \hspace{1cm} [-0.1163, 0.1254] & \hspace{1cm} 0.1208 \\
3                 &\hspace{0.8mm}  300  &\hspace{0.8mm}  [-0.0447, 0.0539]    & \hspace{1cm}   0.0491  &  \hspace{1cm} [-0.1120, 0.1211] & \hspace{1cm} 0.1164 \\
3                 &\hspace{0.8mm}  500  &\hspace{0.8mm}  [-0.0389, 0.0480]    & \hspace{1cm}   0.0432  &  \hspace{1cm} [-0.1116, 0.1207] & \hspace{1cm} 0.1161 \\
3                 &\hspace{0.8mm}  1000  &\hspace{0.8mm} [-0.0320, 0.0412]    & \hspace{1cm}   0.0364  &  \hspace{1cm} [-0.1113, 0.1204] & \hspace{1cm} 0.1158 \\
3                 &\hspace{0.8mm}  1500  &\hspace{0.8mm} [-0.0286, 0.0377]    & \hspace{1cm}   0.0329  &  \hspace{1cm} [-0.1113, 0.1203] & \hspace{1cm} 0.1157 \\
3                 &\hspace{0.8mm}  2000  &\hspace{0.8mm} [-0.0234, 0.0325]    & \hspace{1cm}   0.0277  &  \hspace{1cm} [-0.1112, 0.1203] & \hspace{1cm} 0.1156 \\
\hline\hline
\multicolumn{6}{c}{ $3\sigma$ C.L. \hspace{3cm}  $\delta_{sys}=0\%$  \hspace{4cm}  $\delta_{sys}=5\%$ }\\
\hline
1.4               &\hspace{0.8mm}  50   &\hspace{0.8mm}  [-0.1209, 0.1683]    & \hspace{1cm}   0.1762  &   \hspace{1cm} [-0.2096, 0.2599] & \hspace{1cm} 0.2567 \\
1.4               &\hspace{0.8mm}  300   &\hspace{0.8mm} [-0.0703, 0.1177]    & \hspace{1cm}   0.1126  &   \hspace{1cm} [-0.1968, 0.2472] & \hspace{1cm} 0.2438 \\
1.4               &\hspace{0.8mm}  500  &\hspace{0.8mm}  [-0.0598, 0.1071]    & \hspace{1cm}   0.0990  &   \hspace{1cm} [-0.1957, 0.2460] & \hspace{1cm} 0.2427 \\
1.4               &\hspace{0.8mm}  1000  &\hspace{0.8mm} [-0.0477, 0.0950]    & \hspace{1cm}   0.0833  &   \hspace{1cm} [-0.1948, 0.2452] & \hspace{1cm} 0.2418 \\
1.4               &\hspace{0.8mm}  1500  &\hspace{0.8mm} [-0.0416, 0.0889]    & \hspace{1cm}   0.0752  &   \hspace{1cm} [-0.1945, 0.2449] & \hspace{1cm} 0.2416 \\
\hline
3                 &\hspace{0.8mm}  50   &\hspace{0.8mm}  [-0.0794, 0.0886]    & \hspace{1cm}   0.0838  &   \hspace{1cm} [-0.1273, 0.1364] & \hspace{1cm} 0.1318 \\
3                 &\hspace{0.8mm}  300  &\hspace{0.8mm}  [-0.0492, 0.0583]    & \hspace{1cm}   0.0535  &   \hspace{1cm} [-0.1226, 0.1317] & \hspace{1cm} 0.1270 \\
3                 &\hspace{0.8mm}  500  &\hspace{0.8mm}  [-0.0428, 0.0520]    & \hspace{1cm}   0.0472  &   \hspace{1cm} [-0.1222, 0.1313] & \hspace{1cm} 0.1266 \\
3                 &\hspace{0.8mm}  1000  &\hspace{0.8mm} [-0.0353, 0.0445]    & \hspace{1cm}   0.0396  &   \hspace{1cm} [-0.1219, 0.1310] & \hspace{1cm} 0.1263 \\
3                 &\hspace{0.8mm}  1500  &\hspace{0.8mm} [-0.0315, 0.0407]    & \hspace{1cm}   0.0358  &   \hspace{1cm} [-0.1218, 0.1309] & \hspace{1cm} 0.1262 \\
3                 &\hspace{0.8mm}  2000  &\hspace{0.8mm} [-0.0258, 0.0350]    & \hspace{1cm}   0.0301  &   \hspace{1cm} [-0.1217, 0.1308] & \hspace{1cm} 0.1261 \\
\hline\hline
\end{tabular}
\end{center}
\end{table}

In Fig. 4 we used two center-of-mass energies $\sqrt{s}=1.4, 3\hspace{0.8mm}TeV$ expected for the CLIC
accelerator in order to get contour limits in the plane $\hat a_V-\hat a_A$ for $\gamma e^- \to \bar t b \nu_e$
and the expected luminosities of ${\cal L}=50, 500, 1500, 2000\hspace{0.8mm}fb^{-1}$ with polarized and unpolarized beams
of electrons and positrons.

As an indicator of the order of magnitude, using $b$-tagging efficiency of 0.8 and considering the
systematic errors of $\delta_{sys}=0\%,\hspace{1mm}5\%$, in Tables I and II we present the bounds obtained on
the $\hat a_V$ magnetic moment and $\hat a_A$ electric dipole moments of the t-quark
with the polarization $P_{e^-}=-80\%$ for the electron beams, $P_{e^+}=0\%$ for the positron, as well as with
unpolarized beams, with $\sqrt{s}= 1.4, 3$\hspace{0.8mm}$TeV$, ${\cal L} = 50, 300, 500,
1000, 1500, 2000$\hspace{0.8mm}$fb^{-1}$ at $2\sigma$ and $3\sigma$ $C.L.$, respectively.
As expected, the results presented in Tables I and II clearly show that as the energy and luminosity of the
collider increase, the bounds on the dipole moments of the top quark are stronger.
We observed that these results are competitive with those recently reported in previous
studies \cite{Baur,Bouzas,Bouzas1,Sh}. From results presented in Table I, it is obvious that the effect
of polarized beams is more significant than the effect of unpolarized beams (see Table II).

\begin{table}[!ht]
\caption{Bounds on the $\hat a_V$ magnetic moment and $\hat a_A$ electric dipole moment for the process $\gamma e^- \to \bar t b \nu_e$
($\gamma$ is the Compton backscattering photon) for $b-\mbox{tagging efficiency}=0.8$, $\delta_{sys}=0\%, \hspace{2mm} 5\%$ at $2\sigma$
and $3\sigma$ C.L.}
\begin{center}
\begin{tabular}{cccccc}
\hline\hline
\multicolumn{6}{c}{ $2\sigma$ C.L. \hspace{3cm}  $\delta_{sys}=0\%$  \hspace{4cm}  $\delta_{sys}=5\%$ }\\
\hline
\cline{1-6} $\sqrt{s}\hspace{0.8mm}(TeV)$  & \hspace{0.8mm} ${\cal L}\hspace{0.8mm}(fb^{-1})$  & \hspace{1cm} $\hat a_V$   & $\hspace{1cm} |\hat a_A|$
& \hspace{1cm} $\hat a_V$   & $\hspace{1cm} |\hat a_A|$ \\
\hline
1.4               &\hspace{8mm}  50   &\hspace{8mm}  [-0.1624, 0.2158]    & \hspace{1cm}   0.1872  &  \hspace{8mm} [-0.2198, 0.2733] & \hspace{1cm} 0.2451 \\
1.4               &\hspace{8mm}  300   &\hspace{8mm} [-0.0958, 0.1493]    & \hspace{1cm}   0.1196  &  \hspace{8mm} [-0.2003, 0.2538] & \hspace{1cm} 0.2255 \\
1.4               &\hspace{8mm}  500  &\hspace{8mm}  [-0.0882, 0.1353]    & \hspace{1cm}   0.1052  &  \hspace{8mm} [-0.1985, 0.2520] & \hspace{1cm} 0.2237 \\
1.4               &\hspace{8mm}  1000  &\hspace{8mm} [-0.0657, 0.1192]    & \hspace{1cm}   0.0885  &  \hspace{8mm} [-0.1971, 0.2506] & \hspace{1cm} 0.2223 \\
1.4               &\hspace{8mm}  1500  &\hspace{8mm} [-0.0576, 0.1111]    & \hspace{1cm}   0.0800  &  \hspace{8mm} [-0.1967, 0.2501] & \hspace{1cm} 0.2218 \\
\hline
3                 &\hspace{8mm}  50   &\hspace{8mm}  [-0.0845, 0.0936]    & \hspace{1cm}   0.0887  &  \hspace{8mm} [-0.1201, 0.1292] & \hspace{1cm} 0.1246 \\
3                 &\hspace{8mm}  300  &\hspace{8mm}  [-0.0523, 0.0614]    & \hspace{1cm}   0.0563  &  \hspace{8mm} [-0.1127, 0.1218] & \hspace{1cm} 0.1172 \\
3                 &\hspace{8mm}  500  &\hspace{8mm}  [-0.0454, 0.0545]    & \hspace{1cm}   0.0494  &  \hspace{8mm} [-0.1120, 0.1212] & \hspace{1cm} 0.1165 \\
3                 &\hspace{8mm}  1000  &\hspace{8mm} [-0.0375, 0.0465]    & \hspace{1cm}   0.0414  &  \hspace{8mm} [-0.1115, 0.1207] & \hspace{1cm} 0.1160 \\
3                 &\hspace{8mm}  1500  &\hspace{8mm} [-0.0334, 0.0425]    & \hspace{1cm}   0.0373  &  \hspace{8mm} [-0.1114, 0.1205] & \hspace{1cm} 0.1159 \\
3                 &\hspace{8mm}  2000  &\hspace{8mm} [-0.0273, 0.0364]    & \hspace{1cm}   0.0312  &  \hspace{8mm} [-0.1113, 0.1204] & \hspace{1cm} 0.1158 \\
\hline\hline
\multicolumn{6}{c}{ $3\sigma$ C.L. \hspace{4.2cm}  $\delta_{sys}=0\%$  \hspace{4.2cm}  $\delta_{sys}=5\%$ }\\
\hline
1.4               &\hspace{8mm}  50   &\hspace{8mm}  [-0.1793, 0.2327]    & \hspace{1cm}   0.2043  &  \hspace{8mm} [-0.2420, 0.2955] & \hspace{1cm} 0.2674 \\
1.4               &\hspace{8mm}  300   &\hspace{8mm} [-0.1065, 0.1599]    & \hspace{1cm}   0.1305  &  \hspace{8mm} [-0.2207, 0.2742] & \hspace{1cm} 0.2460 \\
1.4               &\hspace{8mm}  500  &\hspace{8mm}  [-0.0912, 0.1447]    & \hspace{1cm}   0.1149  &  \hspace{8mm} [-0.2188, 0.2722] & \hspace{1cm} 0.2440 \\
1.4               &\hspace{8mm}  1000  &\hspace{8mm} [-0.0735, 0.1269]    & \hspace{1cm}   0.0969  &  \hspace{8mm} [-0.2172, 0.2707] & \hspace{1cm} 0.2425 \\
1.4               &\hspace{8mm}  1500  &\hspace{8mm} [-0.0645, 0.1180]    & \hspace{1cm}   0.0873  &  \hspace{8mm} [-0.2167, 0.2702] & \hspace{1cm} 0.2420 \\
\hline
3                 &\hspace{8mm}  50   &\hspace{8mm}  [-0.0926, 0.1017]    & \hspace{1cm}   0.0966  &  \hspace{8mm} [-0.1315, 0.1406] & \hspace{1cm} 0.1360 \\
3                 &\hspace{8mm}  300  &\hspace{8mm}  [-0.0574, 0.0665]    & \hspace{1cm}   0.0616  &  \hspace{8mm} [-0.1234, 0.1325] & \hspace{1cm} 0.1279 \\
3                 &\hspace{8mm}  500  &\hspace{8mm}  [-0.0500, 0.0591]    & \hspace{1cm}   0.0541  &  \hspace{8mm} [-0.1226, 0.1318] & \hspace{1cm} 0.1271 \\
3                 &\hspace{8mm}  1000  &\hspace{8mm} [-0.0413, 0.0504]    & \hspace{1cm}   0.0453  &  \hspace{8mm} [-0.1221, 0.1312] & \hspace{1cm} 0.1266 \\
3                 &\hspace{8mm}  1500  &\hspace{8mm} [-0.0368, 0.0459]    & \hspace{1cm}   0.0408  &  \hspace{8mm} [-0.1219, 0.1310] & \hspace{1cm} 0.1264 \\
3                 &\hspace{8mm}  2000  &\hspace{8mm} [-0.0301, 0.0392]    & \hspace{1cm}   0.0341  &  \hspace{8mm} [-0.1218, 0.1309] & \hspace{1cm} 0.1263 \\
\hline\hline
\end{tabular}
\end{center}
\end{table}

To complement our results, in Table III we show the single top production total cross section for the process
$\gamma e^- \to \bar t b \nu_e$ as a function of the dipole moments $\hat a_V$ and $\hat a_A$ at the two CLIC energies
of 1.4 and 3$\hspace{0.8mm}TeV$. For polarized beams ($P_{e^-, e^+}=-80\%, 0\%$), the total cross section most
significant for the process considered is $\sigma_{Tot}(\gamma e^-\to \bar t b\nu_e)=5.4330\times 10^{-1}\hspace{0.8mm}pb$
for $\hat a_V=-1$, $\hat a_A=0$ and $\sqrt{s}=3\hspace{0.8mm}TeV$, while for $\hat a_A=1$, $\hat a_V=0$ and $\sqrt{s}=3\hspace{0.8mm}TeV$
the total cross section is $\sigma_{Tot}(\gamma e^-\to \bar t b\nu_e)=5.3923\times 10^{-1}\hspace{0.8mm}pb$. On the other
hand, for unpolarized beams ($P_{e^-, e^+}=0\%, 0\%$), the total cross sections are $\sigma_{Tot}(\gamma e^-\to \bar t b\nu_e)=3.0186\times 10^{-1}\hspace{0.8mm}pb$ for $\hat a_V=-1$, $\hat a_A=0$ and $\sigma_{Tot}(\gamma e^-\to \bar t b\nu_e)=2.9953\times 10^{-1}\hspace{0.8mm}pb$ for $\hat a_V=0$, $\hat a_A=-1$ with $\sqrt{s}=3\hspace{0.8mm}TeV$, respectively. Therefore, the total cross section for the case of polarized beams
shows improvement by a factor of 1.8 with respect to the unpolarized case.

\begin{table}[!ht]
\caption{Total cross sections for the process $\gamma e^- \to \bar t b \nu_e$ ($\gamma$ is the Compton backscattering photon) as a function
of $\hat a_V$ and $\hat a_A$ for $b-\mbox{tagging efficiency}=0.8$ at $P_{e^-, e^+}=-80\%, 0\%$ and $P_{e^-, e^+}=0\%, 0\%$, respectively.}
\begin{center}
\begin{tabular}{c|cccc}
\hline\hline
\multicolumn{5}{c}{             \hspace{3.5cm}  $P_{e^-, e^+}=-80\%, 0\%$  \hspace{3cm} }\\
\hline
\cline{1-5} $\sqrt{s}\hspace{0.8mm}(TeV)$  & \hspace{0.8mm} ${\hat a_V}$  & \hspace{1cm} $\sigma_{Tot}(\gamma e^-\to \bar t b\nu_e)(pb)$   & $\hspace{1cm} \hat a_A$
& \hspace{1cm} $\sigma_{Tot}(\gamma e^-\to \bar t b\nu_e)(pb)$ \\
\hline
1.4               &\hspace{0.8mm}  -1     &\hspace{0.8mm} $1.4328\times 10^{-1}$    & \hspace{1cm}    -1    &  \hspace{1cm} 1.3862$\times 10^{-1}$  \\
1.4               &\hspace{0.8mm}  -0.5   &\hspace{0.8mm} $7.5675\times 10^{-2}$     & \hspace{1cm}   -0.5  &  \hspace{1cm} 7.3354$\times 10^{-2}$ \\
1.4               &\hspace{0.8mm}   0     &\hspace{1.1cm} $5.1582\times 10^{-2}$ (SM)    & \hspace{1cm}    0    &  \hspace{2cm} 5.1582$\times 10^{-2}$ (SM)  \\
1.4               &\hspace{0.8mm}   0.5   &\hspace{0.8mm} $7.1010\times 10^{-2}$     & \hspace{1cm}    0.5  &  \hspace{1cm} 7.3332$\times 10^{-2}$ \\
1.4               &\hspace{0.8mm}   1     &\hspace{0.8mm} $1.3393\times 10^{-1}$     & \hspace{1cm}    1    &  \hspace{1cm} 1.3862$\times 10^{-1}$  \\
\hline
3               &\hspace{0.8mm}    -1     &\hspace{0.8mm} $5.4330\times 10^{-1}$    & \hspace{1cm}    -1    &  \hspace{1cm} 5.3912$\times 10^{-1}$  \\
3               &\hspace{0.8mm}    -0.5   &\hspace{0.8mm} $1.9331\times 10^{-1}$     & \hspace{1cm}   -0.5  &  \hspace{1cm} 1.9120$\times 10^{-1}$ \\
3               &\hspace{0.8mm}     0     &\hspace{1.1cm} $7.5185\times 10^{-2}$ (SM)    & \hspace{1cm}    0    &  \hspace{2cm} 7.5185$\times 10^{-2}$ (SM) \\
3               &\hspace{0.8mm}     0.5   &\hspace{0.8mm} $1.8907\times 10^{-1}$     & \hspace{1cm}    0.5  &  \hspace{1cm} 1.9117$\times 10^{-1}$ \\
3               &\hspace{0.8mm}     1     &\hspace{0.8mm} $5.3492\times 10^{-1}$     & \hspace{1cm}    1    &  \hspace{1cm} 5.3923$\times 10^{-1}$  \\
\hline\hline
\multicolumn{5}{c}{  \hspace{3.5cm}  $P_{e^-, e^+}=0\%, 0\%$  \hspace{3cm}   }\\
\hline
1.4               &\hspace{0.8mm}  -1     &\hspace{0.8mm} $7.9603\times 10^{-2}$    & \hspace{1cm}    -1    &  \hspace{1cm} 7.7002$\times 10^{-2}$  \\
1.4               &\hspace{0.8mm}  -0.5   &\hspace{0.8mm} $4.2039\times 10^{-2}$     & \hspace{1cm}   -0.5  &  \hspace{1cm} 4.0744$\times 10^{-2}$ \\
1.4               &\hspace{0.8mm}   0     &\hspace{1.1cm} $2.8659\times 10^{-2}$ (SM)    & \hspace{1cm}    0    &  \hspace{2cm} 2.8659$\times 10^{-2}$ (SM)  \\
1.4               &\hspace{0.8mm}   0.5   &\hspace{0.8mm} $3.9456\times 10^{-2}$     & \hspace{1cm}    0.5  &  \hspace{1cm} 4.0742$\times 10^{-2}$ \\
1.4               &\hspace{0.8mm}   1     &\hspace{0.8mm} $7.4405\times 10^{-2}$     & \hspace{1cm}    1    &  \hspace{1cm} 7.7019$\times 10^{-2}$  \\
\hline
3               &\hspace{0.8mm}    -1     &\hspace{0.8mm} $3.0186\times 10^{-1}$    & \hspace{1cm}    -1    &  \hspace{1cm} 2.9953$\times 10^{-1}$  \\
3               &\hspace{0.8mm}    -0.5   &\hspace{0.8mm} $1.0738\times 10^{-1}$     & \hspace{1cm}   -0.5  &  \hspace{1cm} 1.0621$\times 10^{-1}$ \\
3               &\hspace{0.8mm}     0     &\hspace{1.1cm} $4.1765\times 10^{-2}$ (SM)    & \hspace{1cm}    0    &  \hspace{2cm} 4.1765$\times 10^{-2}$ (SM)  \\
3               &\hspace{0.8mm}     0.5   &\hspace{0.8mm} $1.0502\times 10^{-1}$     & \hspace{1cm}    0.5  &  \hspace{1cm} 1.0620$\times 10^{-1}$ \\
3               &\hspace{0.8mm}     1     &\hspace{0.8mm} $2.9713\times 10^{-1}$     & \hspace{1cm}    1    &  \hspace{1cm} 2.9952$\times 10^{-1}$  \\
\hline\hline
\end{tabular}
\end{center}
\end{table}

\section{Weizsacker-Williams Approximation (WWA): Cross section of $e^+e^- \to e^+\gamma^* e^- \to \bar t b \nu_e e^+$}

We use the WWA and consider the process $e^+e^- \to e^+\gamma^* e^- \to \bar t b \nu_e e^+$ which is potentially
useful for studying the dipole moments of the top quark with polarized and unpolarized $e^-$ beams and for the center-of-mass
energies of the CLIC \cite{CLIC}.

\subsection{Top quark dipole moments through the process $e^+e^- \to e^+\gamma^* e^- \to \bar t b \nu_e e^+$
with polarized and unpolarized beams}

The Feynman diagrams for the subprocess $\gamma^* e^- \to \bar t b \nu_e $ are shown in Fig. 2.
The total cross section of the subprocess depends on the contribution of the SM,
(diagrams (1)-(3)) plus the contribution of the anomalous couplings (diagram (4)).

For the study of the process $e^+e^- \to e^+\gamma^* e^- \to \bar t b \nu_e e^+$, in Fig. 5 we show the
total cross section as a function of the electromagnetic form factors of the top quark $\hat a_V$ and
$\hat a_A$ for $P_{e^-, e^+}=-80\%, 0\%$ \cite{Moortgat}, two different center-of-mass energies
$\sqrt{s}=1.4, 3$\hspace{0.8mm}$TeV$ \cite{CLIC} and the Weizsacker-Williams photon virtuality
$Q^2 = 2$\hspace{0.8mm}$GeV^2$ \cite{L3,OPAL,Sahin,Billur}. We can see from this figure that the total
cross section changes strongly with $\sqrt{s}$ reaching $20\%$ and $23\%$ at the end of the range considered
to $\hat a_{V, A}$ with polarized and unpolarized beams.

\begin{table}[!ht]
\caption{Bounds on the $\hat a_V$ magnetic moment and $\hat a_A$ electric dipole moment for the process $e^+e^- \to e^+\gamma^* e^- \to \bar t b \nu_e e^+$ ($\gamma^*$ is the Weizsacker-Williams photon) for $Q^2 = 2$\hspace{0.8mm}$GeV^2$, $P_{e^-, e^+}=-80\%, 0\%$, $b-\mbox{tagging efficiency}=0.8$, $\delta_{sys}=0\%, \hspace{2mm} 5\%$ at $2\sigma$ and $3\sigma$ C.L.}
\begin{center}
\begin{tabular}{cccccc}
\hline\hline
\multicolumn{6}{c}{ $2\sigma$ C.L. \hspace{3cm}  $\delta_{sys}=0\%$  \hspace{4cm}  $\delta_{sys}=5\%$ }\\
\hline
\cline{1-6} $\sqrt{s}\hspace{0.8mm}(TeV)$  & \hspace{0.8mm} ${\cal L}\hspace{0.8mm}(fb^{-1})$  & \hspace{1cm} $\hat a_V$   & $\hspace{1cm} |\hat a_A|$
& \hspace{1cm} $\hat a_V$   & $\hspace{1cm} |\hat a_A|$ \\
\hline
1.4              &\hspace{8mm}  50   &\hspace{8mm}  [-0.3474, 0.4397]    & \hspace{1cm}   0.3908  &  \hspace{8mm} [-0.3669, 0.4592] & \hspace{1cm} 0.4104 \\
1.4              &\hspace{8mm}  300   &\hspace{8mm} [-0.2078, 0.3001]    & \hspace{1cm}   0.2497  &  \hspace{8mm} [-0.2647, 0.3570] & \hspace{1cm} 0.3074 \\
1.4              &\hspace{8mm}  500  &\hspace{8mm}  [-0.1784, 0.2707]    & \hspace{1cm}   0.2197  &  \hspace{8mm} [-0.2505, 0.3428] & \hspace{1cm} 0.2931 \\
1.4              &\hspace{8mm}  1000  &\hspace{8mm} [-0.1443, 0.2366]    & \hspace{1cm}   0.1848  &  \hspace{8mm} [-0.2383, 0.3306] & \hspace{1cm} 0.2807 \\
1.4              &\hspace{8mm}  1500  &\hspace{8mm} [-0.1271, 0.2194]    & \hspace{1cm}   0.1670  &  \hspace{8mm} [-0.2339, 0.3262] & \hspace{1cm} 0.2762 \\
\hline
3                &\hspace{8mm}  50   &\hspace{8mm}  [-0.1806, 0.2180]    & \hspace{1cm}   0.1984  &  \hspace{8mm} [-0.2037, 0.2411] & \hspace{1cm} 0.2216 \\
3                &\hspace{8mm}  300  &\hspace{8mm}  [-0.1094, 0.1469]    & \hspace{1cm}   0.1267  &  \hspace{8mm} [-0.1652, 0.2026] & \hspace{1cm} 0.1829 \\
3                &\hspace{8mm}  500  &\hspace{8mm}  [-0.0944, 0.1318]    & \hspace{1cm}   0.1115  &  \hspace{8mm} [-0.1608, 0.1982] & \hspace{1cm} 0.1785 \\
3                &\hspace{8mm}  1000  &\hspace{8mm} [-0.0769, 0.1144]    & \hspace{1cm}   0.0937  &  \hspace{8mm} [-0.1573, 0.1947] & \hspace{1cm} 0.1750 \\
3                &\hspace{8mm}  1500  &\hspace{8mm} [-0.0681, 0.1055]    & \hspace{1cm}   0.0847  &  \hspace{8mm} [-0.1561, 0.1935] & \hspace{1cm} 0.1738 \\
3                &\hspace{8mm}  2000  &\hspace{8mm} [-0.0547, 0.0921]    & \hspace{1cm}   0.0713  &  \hspace{8mm} [-0.1555, 0.1929] & \hspace{1cm} 0.1732 \\
\hline\hline
\multicolumn{6}{c}{ $3\sigma$ C.L. \hspace{4.2cm}  $\delta_{sys}=0\%$  \hspace{4.2cm}  $\delta_{sys}=5\%$ }\\
\hline
1.4              &\hspace{8mm}  50   &\hspace{8mm}  [-0.3827, 0.4750]    & \hspace{1cm}   0.4264  &  \hspace{8mm} [-0.4040, 0.4963] & \hspace{1cm} 0.4478 \\
1.4              &\hspace{8mm}  300   &\hspace{8mm} [-0.2302, 0.3225]    & \hspace{1cm}   0.2724  &  \hspace{8mm} [-0.2924, 0.3847] & \hspace{1cm} 0.3354 \\
1.4              &\hspace{8mm}  500  &\hspace{8mm}  [-0.1980, 0.2903]    & \hspace{1cm}   0.2397  &  \hspace{8mm} [-0.2769, 0.3692] & \hspace{1cm} 0.3197 \\
1.4              &\hspace{8mm}  1000  &\hspace{8mm} [-0.1607, 0.2530]    & \hspace{1cm}   0.2016  &  \hspace{8mm} [-0.2636, 0.3559] & \hspace{1cm} 0.3063 \\
1.4              &\hspace{8mm}  1500  &\hspace{8mm} [-0.1418, 0.2341]    & \hspace{1cm}   0.1822  &  \hspace{8mm} [-0.2587, 0.3510] & \hspace{1cm} 0.3013 \\
\hline
3                &\hspace{8mm}  50   &\hspace{8mm}  [-0.1986, 0.2360]    & \hspace{1cm}   0.2165  &  \hspace{8mm} [-0.2238, 0.2612] & \hspace{1cm} 0.2418 \\
3                &\hspace{8mm}  300  &\hspace{8mm}  [-0.1208, 0.1583]    & \hspace{1cm}   0.1383  &  \hspace{8mm} [-0.1817, 0.2191] & \hspace{1cm} 0.1995 \\
3                &\hspace{8mm}  500  &\hspace{8mm}  [-0.1044, 0.1419]    & \hspace{1cm}   0.1217  &  \hspace{8mm} [-0.1770, 0.2144] & \hspace{1cm} 0.1948 \\
3                &\hspace{8mm}  1000  &\hspace{8mm} [-0.0853, 0.1228]    & \hspace{1cm}   0.1023  &  \hspace{8mm} [-0.1732, 0.2106] & \hspace{1cm} 0.1909 \\
3                &\hspace{8mm}  1500  &\hspace{8mm} [-0.0756, 0.1131]    & \hspace{1cm}   0.0924  &  \hspace{8mm} [-0.1718, 0.2092] & \hspace{1cm} 0.1896 \\
3                &\hspace{8mm}  2000  &\hspace{8mm} [-0.0609, 0.1081]    & \hspace{1cm}   0.0777  &  \hspace{8mm} [-0.1711, 0.2086] & \hspace{1cm} 0.1890 \\
\hline\hline
\end{tabular}
\end{center}
\end{table}

In Fig. 6 we present the limit contours for the dipole moments in the
($\hat a_V-\hat a_A$) plane for the process $e^+e^- \to e^+\gamma^* e^- \to \bar t b \nu_e e^+$.
The curves are for $\sqrt{s}=1.4, 3\hspace{0.8mm}TeV$ and ${\cal L}=50, 500, 1500\hspace{0.8mm}fb^{-1}$.
We have used $Q^2=2\hspace{0.8mm}GeV^2$ and $b-\mbox{tagging efficiency}=0.8$.

We summarize the bounds obtained on the anomalous parameters $\hat a_V$ and $\hat a_A$ for $b-\mbox{tagging efficiency}=0.8$,
systematic uncertainties of $\delta_{sys}=0\%,\hspace{1mm}5\%$, $\sqrt{s}=1.4, 3$\hspace{0.8mm}$TeV$, $Q^2 = 2$\hspace{0.8mm}$GeV^2$,
and ${\cal L} = 50, 300, 500, 1000, 1500, 2000$\hspace{0.8mm}$fb^{-1}$ at $2\sigma$ and $3\sigma$ in Tables IV and V.
The bounds obtained on these parameters with polarized/unpolarized beams are slightly moderate with respect to those obtained
by the process $\gamma e^- \to \bar t b \nu_e$ as shown in Tables I (II) and IV (V), respectively.

\begin{table}[!ht]
\caption{Bounds on the $\hat a_V$ magnetic moment and $\hat a_A$ electric dipole moment for the process $e^+e^- \to e^+\gamma^* e^- \to \bar t b \nu_e e^+$ ($\gamma^*$ is the Weizsacker-Williams photon) for $Q^2 = 2$\hspace{0.8mm}$GeV^2$, $b-\mbox{tagging efficiency}=0.8$, $\delta_{sys}=0\%, \hspace{2mm} 5\%$
at $2\sigma$ and $3\sigma$ C.L.}
\begin{center}
\begin{tabular}{cccccc}
\hline\hline
\multicolumn{6}{c}{ $2\sigma$ C.L. \hspace{3cm}  $\delta_{sys}=0\%$  \hspace{4cm}  $\delta_{sys}=5\%$ }\\
\hline
\cline{1-6} $\sqrt{s}\hspace{0.8mm}(TeV)$  & \hspace{0.8mm} ${\cal L}\hspace{0.8mm}(fb^{-1})$  & \hspace{1cm} $\hat a_V$   & $\hspace{1cm} |\hat a_A|$
& \hspace{1cm} $\hat a_V$   & $\hspace{1cm} |\hat a_A|$ \\
\hline
1.4              &\hspace{8mm}  50   &\hspace{8mm}  [-0.4088, 0.5012]    & \hspace{1cm}   0.4527  &  \hspace{8mm} [-0.4219, 0.5142] & \hspace{1cm} 0.4657 \\
1.4              &\hspace{8mm}  300   &\hspace{8mm} [-0.2467, 0.3390]    & \hspace{1cm}   0.2892  &  \hspace{8mm} [-0.2883, 0.3806] & \hspace{1cm} 0.3313 \\
1.4              &\hspace{8mm}  500  &\hspace{8mm}  [-0.2126, 0.0304]    & \hspace{1cm}   0.2545  &  \hspace{8mm} [-0.2673, 0.3597] & \hspace{1cm} 0.3101 \\
1.4              &\hspace{8mm}  1000  &\hspace{8mm} [-0.1728, 0.2651]    & \hspace{1cm}   0.2140  &  \hspace{8mm} [-0.2482, 0.3405] & \hspace{1cm} 0.2907 \\
1.4              &\hspace{8mm}  1500  &\hspace{8mm} [-0.1527, 0.2450]    & \hspace{1cm}   0.1934  &  \hspace{8mm} [-0.2409, 0.3332] & \hspace{1cm} 0.2833 \\
\hline
3                &\hspace{8mm}  50   &\hspace{8mm}  [-0.2126, 0.2493]    & \hspace{1cm}   0.2299  &  \hspace{8mm} [-0.2274, 0.2654] & \hspace{1cm} 0.2459 \\
3                &\hspace{8mm}  300  &\hspace{8mm}  [-0.1300, 0.1666]    & \hspace{1cm}   0.1468  &  \hspace{8mm} [-0.1726, 0.2106] & \hspace{1cm} 0.1908 \\
3                &\hspace{8mm}  500  &\hspace{8mm}  [-0.1125, 0.1491]    & \hspace{1cm}   0.1293  &  \hspace{8mm} [-0.1656, 0.2035] & \hspace{1cm} 0.1838 \\
3                &\hspace{8mm}  1000  &\hspace{8mm} [-0.0921, 0.1287]    & \hspace{1cm}   0.1086  &  \hspace{8mm} [-0.1597, 0.1977] & \hspace{1cm} 0.1778 \\
3                &\hspace{8mm}  1500  &\hspace{8mm} [-0.0818, 0.1184]    & \hspace{1cm}   0.0982  &  \hspace{8mm} [-0.1576, 0.1956] & \hspace{1cm} 0.1757 \\
3                &\hspace{8mm}  2000  &\hspace{8mm} [-0.0662, 0.1028]    & \hspace{1cm}   0.0826  &  \hspace{8mm} [-0.1565, 0.1945] & \hspace{1cm} 0.1747 \\
\hline\hline
\multicolumn{6}{c}{ $3\sigma$ C.L.  \hspace{4.2cm}  $\delta_{sys}=0\%$  \hspace{4.2cm}  $\delta_{sys}=5\%$ }\\
\hline
1.4              &\hspace{8mm}  50   &\hspace{8mm}  [-0.4499, 0.5422]    & \hspace{1cm}   0.4938  &  \hspace{8mm} [-0.4641, 0.5554] & \hspace{1cm} 0.5081 \\
1.4              &\hspace{8mm}  300   &\hspace{8mm} [-0.2727, 0.3651]    & \hspace{1cm}   0.3155  &  \hspace{8mm} [-0.3182, 0.4105] & \hspace{1cm} 0.3614 \\
1.4              &\hspace{8mm}  500  &\hspace{8mm}  [-0.2354, 0.3277]    & \hspace{1cm}   0.2777  &  \hspace{8mm} [-0.2953, 0.3876] & \hspace{1cm} 0.3383 \\
1.4              &\hspace{8mm}  1000  &\hspace{8mm} [-0.1919, 0.2842]    & \hspace{1cm}   0.2335  &  \hspace{8mm} [-0.2744, 0.3667] & \hspace{1cm} 0.3171 \\
1.4              &\hspace{8mm}  1500  &\hspace{8mm} [-0.1698, 0.2622]    & \hspace{1cm}   0.2111  &  \hspace{8mm} [-0.2664, 0.3587] & \hspace{1cm} 0.3091 \\
\hline
3                &\hspace{8mm}  50   &\hspace{8mm}  [-0.2335, 0.2702]    & \hspace{1cm}   0.2508  &  \hspace{8mm} [-0.2497, 0.2877] & \hspace{1cm} 0.2682 \\
3                &\hspace{8mm}  300  &\hspace{8mm}  [-0.1433, 0.1799]    & \hspace{1cm}   0.1602  &  \hspace{8mm} [-0.1899, 0.2279] & \hspace{1cm} 0.2082 \\
3                &\hspace{8mm}  500  &\hspace{8mm}  [-0.1242, 0.1607]    & \hspace{1cm}   0.1410  &  \hspace{8mm} [-0.1822, 0.2202] & \hspace{1cm} 0.2005 \\
3                &\hspace{8mm}  1000  &\hspace{8mm} [-0.1019, 0.1385]    & \hspace{1cm}   0.1185  &  \hspace{8mm} [-0.1758, 0.2138] & \hspace{1cm} 0.1940 \\
3                &\hspace{8mm}  1500  &\hspace{8mm} [-0.0906, 0.1272]    & \hspace{1cm}   0.1071  &  \hspace{8mm} [-0.1735, 0.2115] & \hspace{1cm} 0.1917 \\
3                &\hspace{8mm}  2000  &\hspace{8mm} [-0.0736, 0.1102]    & \hspace{1cm}   0.0901  &  \hspace{8mm} [-0.1723, 0.2103] & \hspace{1cm} 0.1906 \\
\hline\hline
\end{tabular}
\end{center}
\end{table}

Finally, the  predicted values of the corresponding production total cross sections of the process $e^+e^- \to e^+\gamma^* e^- \to \bar t b \nu_e e^+$
are listed in Table VI as a function of $\hat a_V$ and $\hat a_A$ by assuming the initial electron (positron) beam polarization to be $-80\%$ $(0\%)$
for $Q^2=2\hspace{0.8mm}GeV^2$, $b$-tagging efficiency=0.8 and center-of-mass energies of $CLIC-1.4\hspace{0.8mm}TeV$ and $CLIC-3\hspace{0.8mm}TeV$.

It is worth mentioning that the ratio of the total cross section of the process $\gamma e^- \to \bar t b \nu_e$ ($\gamma$ is the Compton backscattering photon) is generally about 18 times greater than the total cross section of the process $e^+e^- \to e^+\gamma^* e^- \to \bar t b \nu_e e^+$ ($\gamma^*$ is the Weizsacker-Williams photon) and both total cross sections depend strongly on the dipole moments ($\hat a_V$ and $\hat a_A$) and on the center-of-mass energy
$(\sqrt{s})$ of the CLIC.

\begin{table}[!ht]
\caption{Total cross sections for the process $e^+e^- \to e^+\gamma^* e^- \to \bar t b \nu_e e^+$ ($\gamma^*$ is the Weizsacker-Williams photon)
as a function of $\hat a_V$ and $\hat a_A$ for $Q^2 = 2$\hspace{0.8mm}$GeV^2$, $b-\mbox{tagging efficiency}=0.8$ at $P_{e^-, e^+}=-80\%, 0\%$ and
$P_{e^-, e^+}=0\%, 0\%$, respectively.}
\begin{center}
\begin{tabular}{c|cccc}
\hline\hline
\multicolumn{5}{c}{             \hspace{3.5cm}  $P_{e^-, e^+}=-80\%, 0\%$  \hspace{3cm} }\\
\hline
\cline{1-5} $\sqrt{s}\hspace{0.5mm}(TeV)$  & \hspace{0.5mm} ${\hat a_V}$  & \hspace{0.5cm} $\sigma_{Tot}(e^+e^- \to e^+\gamma^* e^- \to \bar t b \nu_e e^+                )(pb)$   & $\hspace{0.5cm} \hat a_A$ & \hspace{0.5cm} $\sigma_{Tot}(e^+e^- \to e^+\gamma^* e^- \to \bar t b \nu_e e^+)(pb)$ \\
\hline
1.4               &\hspace{0.8mm}  -1     &\hspace{0.8mm} $7.2429\times 10^{-3}$    & \hspace{0.5cm}    -1   &  \hspace{1cm} 6.9011$\times 10^{-3}$  \\
1.4               &\hspace{0.8mm}  -0.5   &\hspace{0.8mm} $4.2934\times 10^{-3}$    & \hspace{0.5cm}   -0.5  &  \hspace{1cm} 4.1221$\times 10^{-3}$ \\
1.4               &\hspace{0.8mm}   0     &\hspace{1.1cm} $3.1962\times 10^{-3}$ (SM)    & \hspace{0.5cm}    0    &  \hspace{2cm} 3.1962$\times 10^{-3}$ (SM) \\
1.4               &\hspace{0.8mm}   0.5   &\hspace{0.8mm} $3.9510\times 10^{-3}$    & \hspace{0.5cm}    0.5  &  \hspace{1cm} 4.1219$\times 10^{-3}$ \\
1.4               &\hspace{0.8mm}   1     &\hspace{0.8mm} $6.5593\times 10^{-3}$    & \hspace{0.5cm}    1    &  \hspace{1cm} 6.9011$\times 10^{-3}$  \\
\hline
3               &\hspace{0.8mm}    -1     &\hspace{0.8mm} $3.2064\times 10^{-2}$    & \hspace{0.5cm}    -1   &  \hspace{1cm} 3.1205$\times 10^{-2}$  \\
3               &\hspace{0.8mm}    -0.5   &\hspace{0.8mm} $1.4380\times 10^{-2}$    & \hspace{0.5cm}   -0.5  &  \hspace{1cm} 1.3952$\times 10^{-2}$ \\
3               &\hspace{0.8mm}     0     &\hspace{1.1cm} $8.1986\times 10^{-3}$ (SM)   & \hspace{0.5cm}    0    &  \hspace{2cm} 8.1986$\times 10^{-3}$ (SM) \\
3               &\hspace{0.8mm}     0.5   &\hspace{0.8mm} $1.3521\times 10^{-2}$    & \hspace{0.5cm}    0.5  &  \hspace{1cm} 1.3951$\times 10^{-2}$ \\
3               &\hspace{0.8mm}     1     &\hspace{0.8mm} $3.0346\times 10^{-2}$    & \hspace{0.5cm}    1    &  \hspace{1cm} 3.1208$\times 10^{-2}$  \\
\hline\hline
\multicolumn{5}{c}{  \hspace{3.5cm}  $P_{e^-, e^+}=0\%, 0\%$  \hspace{3cm}   }\\
\hline
1.4               &\hspace{0.8mm}  -1     &\hspace{0.8mm} $4.0240\times 10^{-3}$    & \hspace{0.5cm}    -1   &  \hspace{1cm} 3.8340$\times 10^{-3}$  \\
1.4               &\hspace{0.8mm}  -0.5   &\hspace{0.8mm} $2.3853\times 10^{-3}$    & \hspace{0.5cm}   -0.5  &  \hspace{1cm} 2.2903$\times 10^{-3}$ \\
1.4               &\hspace{0.8mm}   0     &\hspace{1.1cm} $1.7754\times 10^{-3}$ (SM)   & \hspace{0.5cm}    0    &  \hspace{2cm} 1.7754$\times 10^{-3}$ (SM) \\
1.4               &\hspace{0.8mm}   0.5   &\hspace{0.8mm} $2.1950\times 10^{-3}$    & \hspace{0.5cm}    0.5  &  \hspace{1cm} 2.2901$\times 10^{-3}$ \\
1.4               &\hspace{0.8mm}   1     &\hspace{0.8mm} $3.6438\times 10^{-3}$    & \hspace{0.5cm}    1    &  \hspace{1cm} 3.8338$\times 10^{-3}$  \\
\hline
3               &\hspace{0.8mm}    -1     &\hspace{0.8mm} $1.7875\times 10^{-2}$    & \hspace{0.5cm}    -1   &  \hspace{1cm} 1.7336$\times 10^{-2}$  \\
3               &\hspace{0.8mm}    -0.5   &\hspace{0.8mm} $7.9900\times 10^{-3}$    & \hspace{0.5cm}   -0.5  &  \hspace{1cm} 7.7515$\times 10^{-3}$ \\
3               &\hspace{0.8mm}     0     &\hspace{1.1cm} $4.5559\times 10^{-3}$ (SM)   & \hspace{0.5cm}    0    &  \hspace{2cm} 4.5559$\times 10^{-3}$ (SM)  \\
3               &\hspace{0.8mm}     0.5   &\hspace{0.8mm} $7.5127\times 10^{-3}$    & \hspace{0.5cm}    0.5  &  \hspace{1cm} 7.7508$\times 10^{-3}$ \\
3               &\hspace{0.8mm}     1     &\hspace{0.8mm} $1.6860\times 10^{-2}$    & \hspace{0.5cm}    1    &  \hspace{1cm} 1.7337$\times 10^{-2}$  \\
\hline\hline
\end{tabular}
\end{center}
\end{table}

\section{Conclusions}

Although $\gamma e^-$ and $\gamma \gamma$ processes require new detectors, $\gamma^{*} e^-$ and
$\gamma^{*} \gamma^{*}$ are produced spontaneously at linear colliders without any detectors.
These processes will allow the future linear colliders to operate in two different modes,
$\gamma^* e^-$ and $\gamma^*\gamma^*$, opening up the opportunity for a wider search for new physics.
Therefore, the $\gamma^* e^-$ linear collisions represent an excellent opportunity to study
top quark anomalous magnetic moment and electric dipole moment.

We have performed a study of the total cross section of the processes $\gamma e^- \to \bar t b \nu_e$
and $e^+e^- \to e^+\gamma^* e^- \to \bar t b \nu_e e^+$, with polarized and unpolarized electron beams
as a function of the anomalous couplings $\hat a_V$ and $\hat a_A$. We have also investigated anomalous
$\hat a_V$ and $\hat a_A$ couplings for both polarized and unpolarized cases. The general behavior of the
cross sections as a function of $\hat a_V$ and $\hat a_A$ couplings does not change. However, we can see from
our calculations of the polarized and unpolarized cases that polarization increases the cross sections. The main
reason for these results can be seen in Fig. 2. There are four diagrams which contribute to the process and one
of them includes the $t\bar t\gamma$ vertex. This diagram gives the maximum contribution to the total cross section.
For $P_{e^-, e^+} =-80\%, 0\%$, this contribution is dominant due to the structure of the $We^-\nu_e$ vertex.
We can appreciate from these figures that lepton polarization can improve the bounds on the
anomalous couplings. The analysis is shown in Figs. 3 and 5 with Compton backscattering photon and Weizsacker-Williams
photon virtuality of $Q^2=2\hspace{0.8mm}GeV^2$ and $b-\mbox{tagging efficiency}=0.8$. In both processes,
the cross section shows a strong dependence on the anomalous couplings $\hat a_V$ and $\hat a_A$, as well as on
the center-of-mass energy $\sqrt{s}$. This variation of the cross section for $\sqrt{s} = 1.4, 3\hspace{0.8mm}TeV$
is of the order $24.5\%$, $26.6\%$ and $20\%$, $23\%$ for $\gamma e^- \to \bar t b \nu_e$ and
$e^+e^- \to e^+\gamma^* e^- \to \bar t b \nu_e e^+$, respectively.

We also include contour plots for the dipole moments at the $95 \%\hspace{1mm}C.L.$ in the $(\hat a_V-\hat a_A)$
plane for the processes $\gamma e^- \to \bar t b \nu_e$ and $e^+e^- \to e^+\gamma^* e^- \to \bar t b \nu_e e^+$ for $Q^2= 2\hspace{0.8mm}GeV^2$ and $\sqrt{s}=1.4, 3\hspace{0.8mm}TeV$ in Figs. 4 and 6. The contours are consistent with the results obtained in Tables I, II, IV and V. The bounds obtained in these Tables are competitive with those recently reported in the literature \cite{Baur,Bouzas,Bouzas1,Sh} and we can observe a strong correlation between the center-of-mass energy $\sqrt{s}$, integrated luminosity ${\cal L}$ and the dipole moments $\hat a_V$ and $\hat a_A$.

Other promising production modes for studying the cross section and the electromagnetic dipole moments $\hat a_V$
and $\hat a_A$ of the top quark are the processes $\gamma \gamma \to t\bar t$ (Compton backscattering photon),
$\gamma^* \gamma^* \to t\bar t$ (Weizsacker-Williams photon) and $\gamma \gamma^* \to t\bar t$ (Compton backscattering
photon, Weizsacker-Williams photon), respectively. These processes are one of the most important sources
of $t\bar t$ pair production and represent new physics effects at a high-energy and high-luminosity linear electron positron
collider as the CLIC.

In conclusion, we have found that the processes $\gamma e^- \to \bar t b \nu_e$ and
$e^+e^- \to e^+\gamma^* e^- \to \bar t b \nu_e e^+$ in the $\gamma e^-$ and $\gamma^* e^-$ collision modes
at the high energies and luminosities expected at the CLIC can be used as a probe to bound the magnetic moment
$\hat a_V$ and electric dipole moment $\hat a_A$ of the top quark. In particular, using integrated luminosity
$2\hspace{0.8mm}ab^{-1}$, center-of-mass energies of 3\hspace{0.8mm}$TeV$,
$b-\mbox{tagging efficiency}=0.8$ and considering the systematic uncertainty $\delta_{sys}=5\%$, we derive
bounds on the dipole moments of the top quark at $2\sigma$ and $3\sigma$ ($90\%$ and $95\%$) C.L.:
$-0.1112\hspace{1mm}(-0.1217) \leq \hat a_V \leq 0.1203\hspace{1mm}(0.1308)$, $|\hat a_A|= 0.1156\hspace{1mm}(0.1261)$
and $-0.1113\hspace{1mm}(-0.1218) \leq \hat a_V \leq 0.1204\hspace{1mm}(0.1309)$, $|\hat a_A|= 0.1158\hspace{1mm}(0.1263)$
for $\gamma e^- \to \bar t b \nu_e$ with unpolarized and polarized $e^-$ beams. For $e^+e^- \to e^+\gamma^* e^- \to \bar t b \nu_e e^+$
with polarized and unpolarized electron beams, $-0.1555\hspace{1mm}(-0.1711) \leq \hat a_V \leq 0.1929\hspace{1mm}(0.2086)$,
$|\hat a_A|= 0.1732\hspace{1mm}(0.1890)$ and $-0.1565\hspace{1mm}(-0.1723) \leq \hat a_V \leq 0.1945\hspace{1mm}(0.2103)$,
$|\hat a_A|= 0.1747\hspace{1mm}(0.1906)$. These results are competitive with those recently reported in previous studies \cite{Baur,Bouzas,Bouzas1,Sh}.
To our knowledge, our numerical results for the dipole moments of the top quark through the single top production
processes have never been reported in the literature before and could be of relevance for the scientific community.

\vspace{1.5cm}

\begin{center}
{\bf Acknowledgments}
\end{center}

A. G. R. acknowledges support from CONACyT, SNI and PROFOCIE (M\'exico).

\newpage

\begin{figure}[t]
\centerline{\scalebox{0.75}{\includegraphics{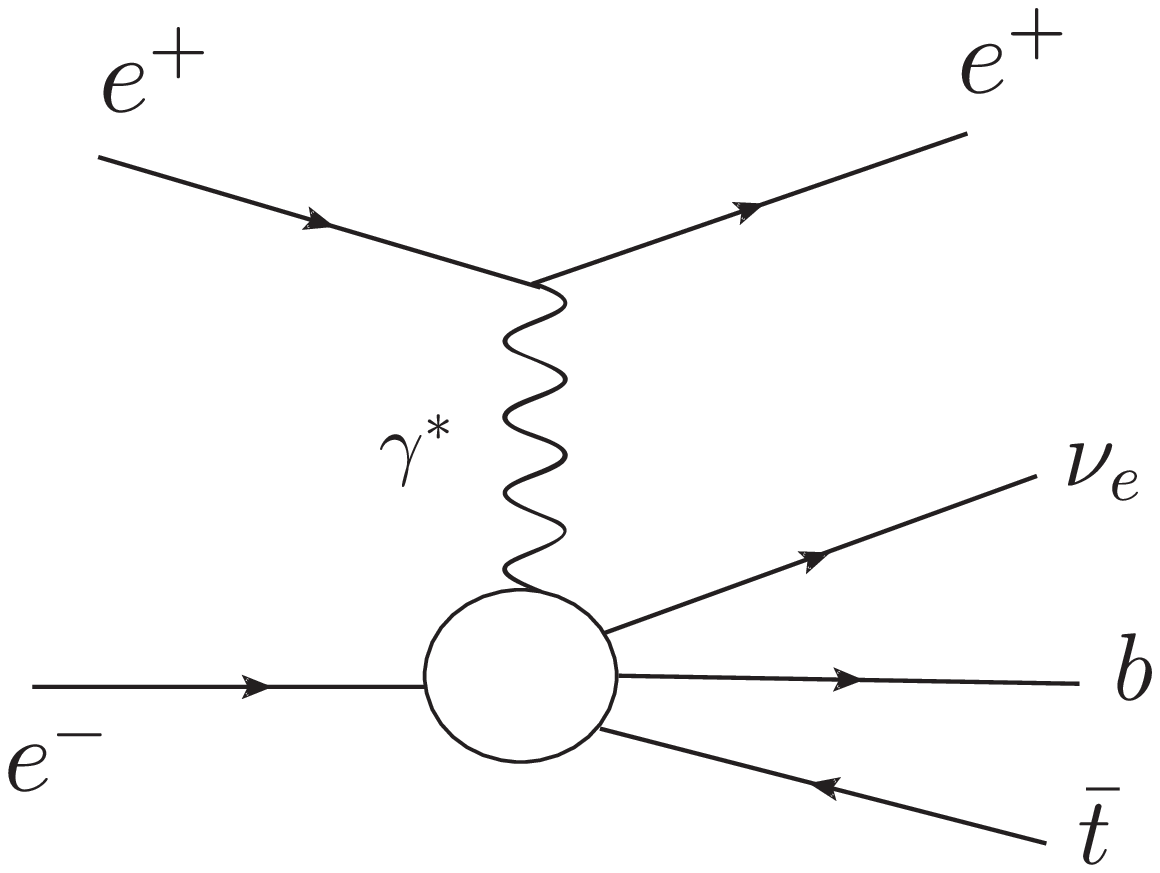}}}
\caption{ \label{fig:gamma} Schematic diagram for the process of single top
quark production $e^+e^- \to e^+\gamma^* e^- \to \bar t b \nu_e e^+ $.}
\end{figure}

\begin{figure}[t]
\centerline{\scalebox{0.75}{\includegraphics{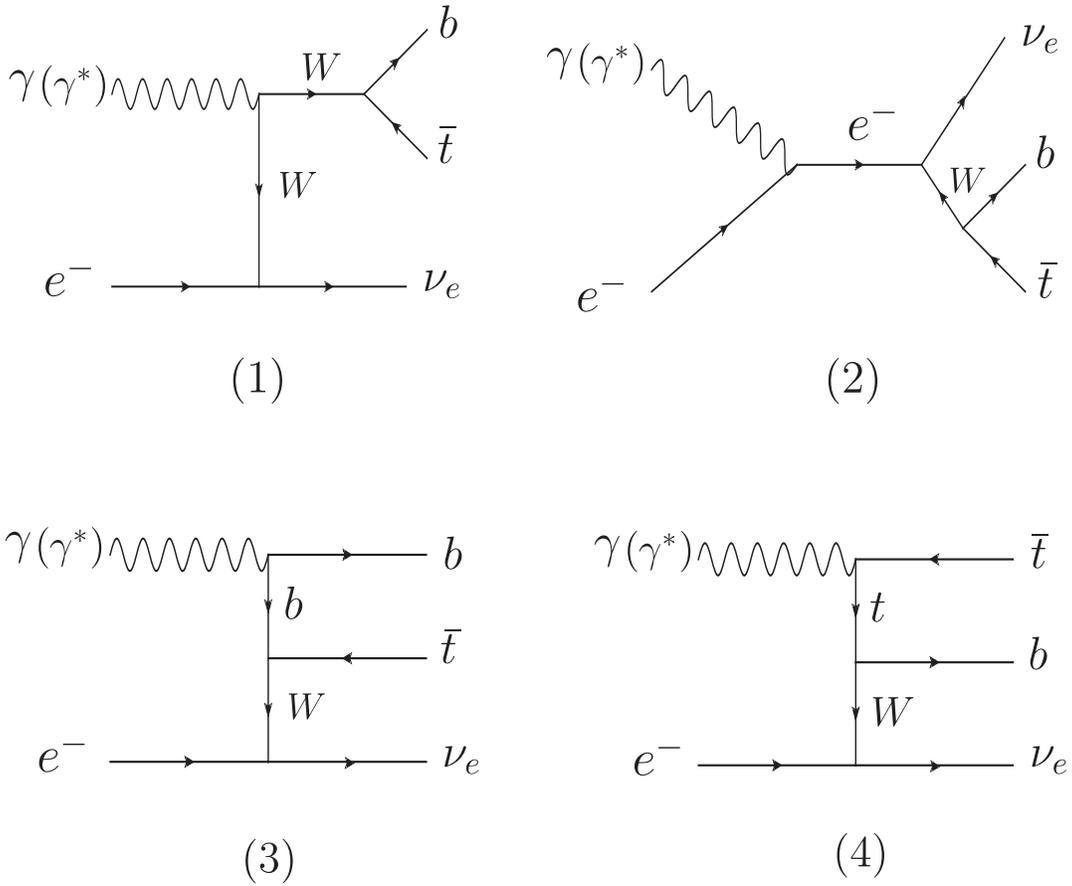}}}
\caption{ \label{fig:gamma} The Feynman diagrams contributing to
the process $\gamma e^- \to \bar t b \nu_e$ and the subprocess
$\gamma^* e^- \to \bar t b \nu_e$.}
\end{figure}

\begin{figure}[t]
\centerline{\scalebox{0.8}{\includegraphics{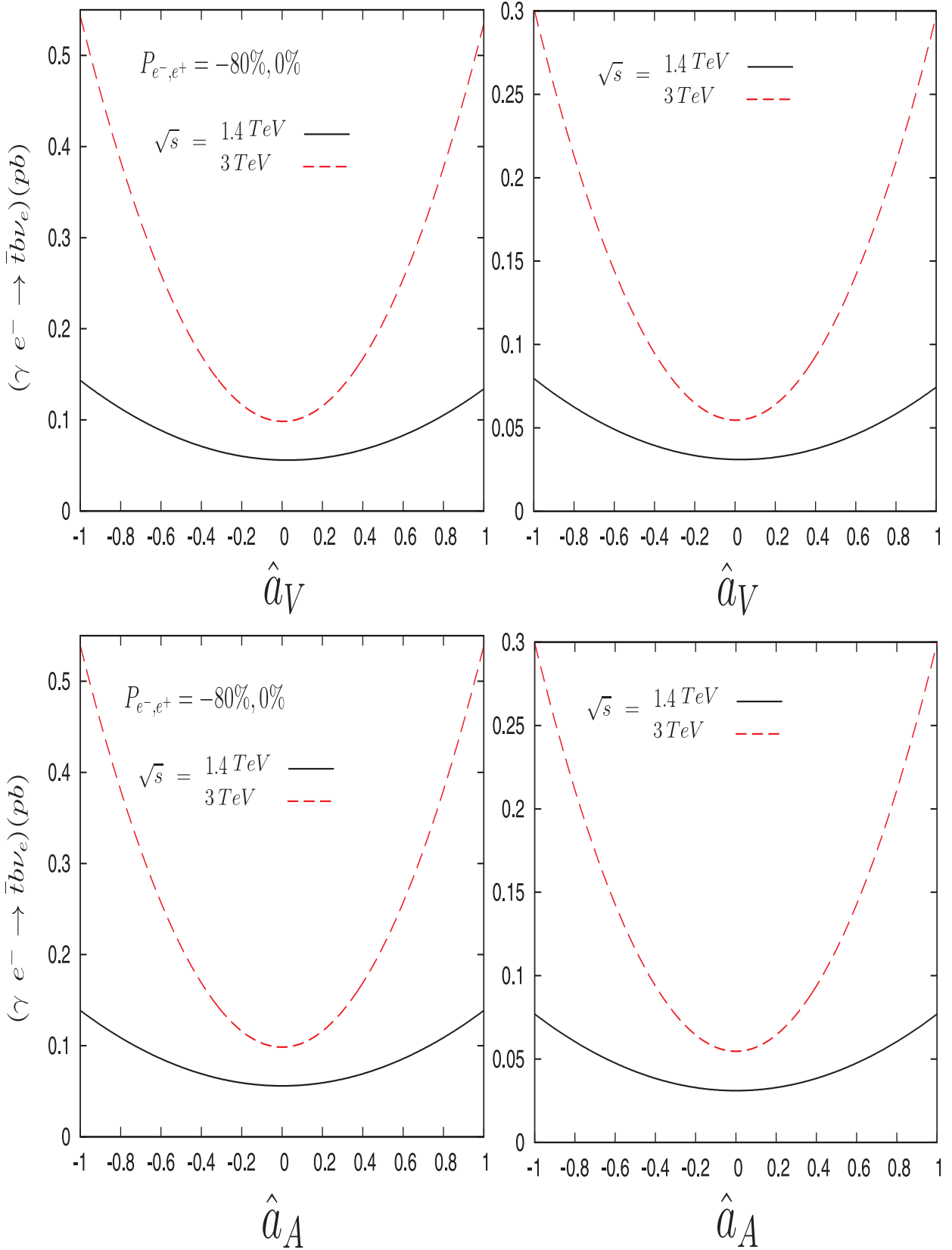}}}
\caption{ \label{fig:gamma1} The integrated total cross section of the process
$\gamma e^- \to \bar t b \nu_e$ ($\gamma$ is the Compton backscattering photon)
as a function of $\hat a_V$ and $\hat a_A$ with $\sqrt{s}=1.4, 3\hspace{0.8mm}TeV$,
and polarized and unpolarized beams.}
\end{figure}

\begin{figure}[t]
\centerline{\scalebox{0.8}{\includegraphics{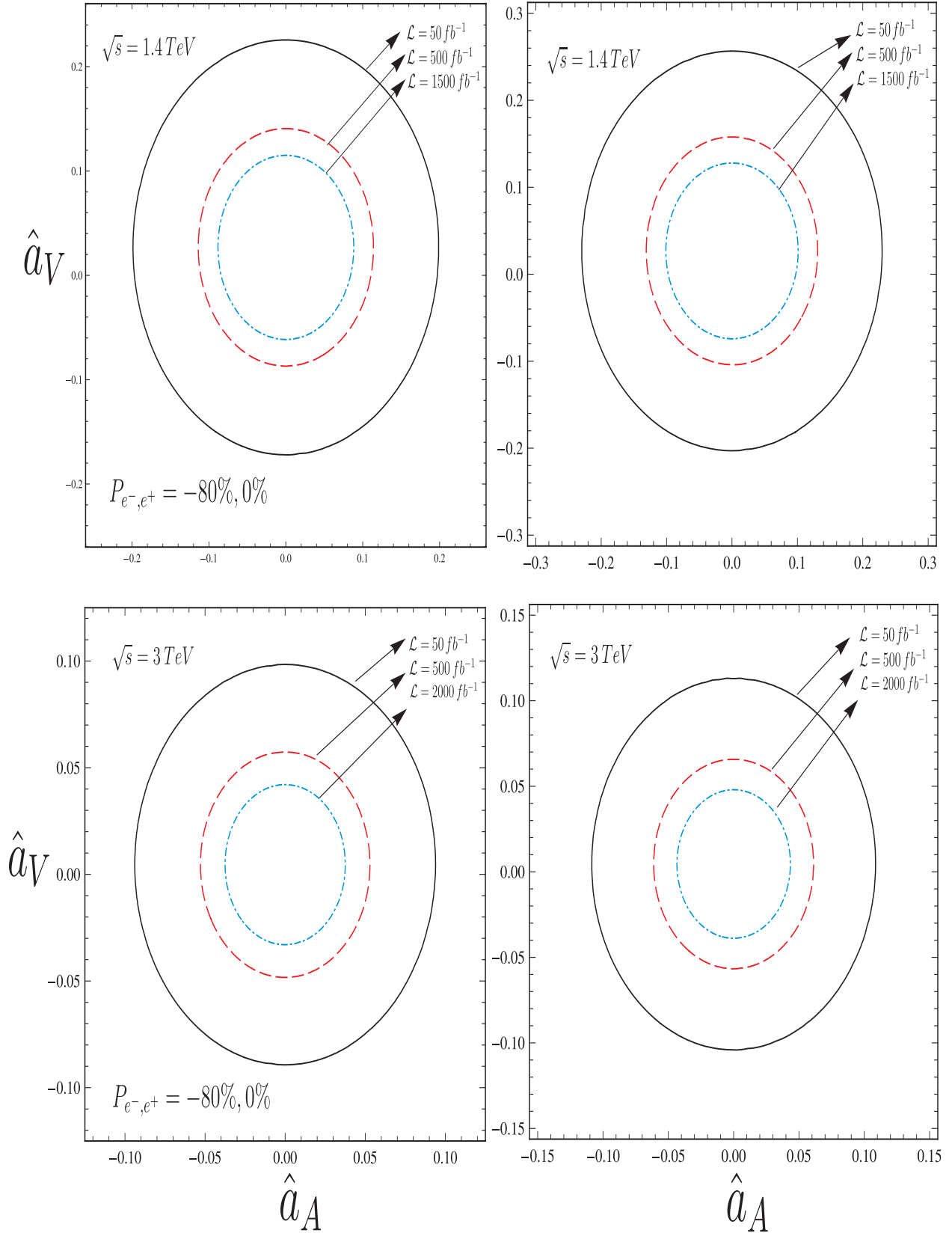}}}
\caption{ \label{fig:gamma2} Limits contours at the $95\% \hspace{1mm}C.L.$ in the
$\hat a_V-\hat a_A$ plane for $\gamma e^- \to \bar t b \nu_e$ ($\gamma$ is the Compton backscattering photon)
and $\sqrt{s}=1.4, 3\hspace{0.8mm}TeV$ with polarized and unpolarized beams.}
\end{figure}

\begin{figure}[t]
\centerline{\scalebox{0.8}{\includegraphics{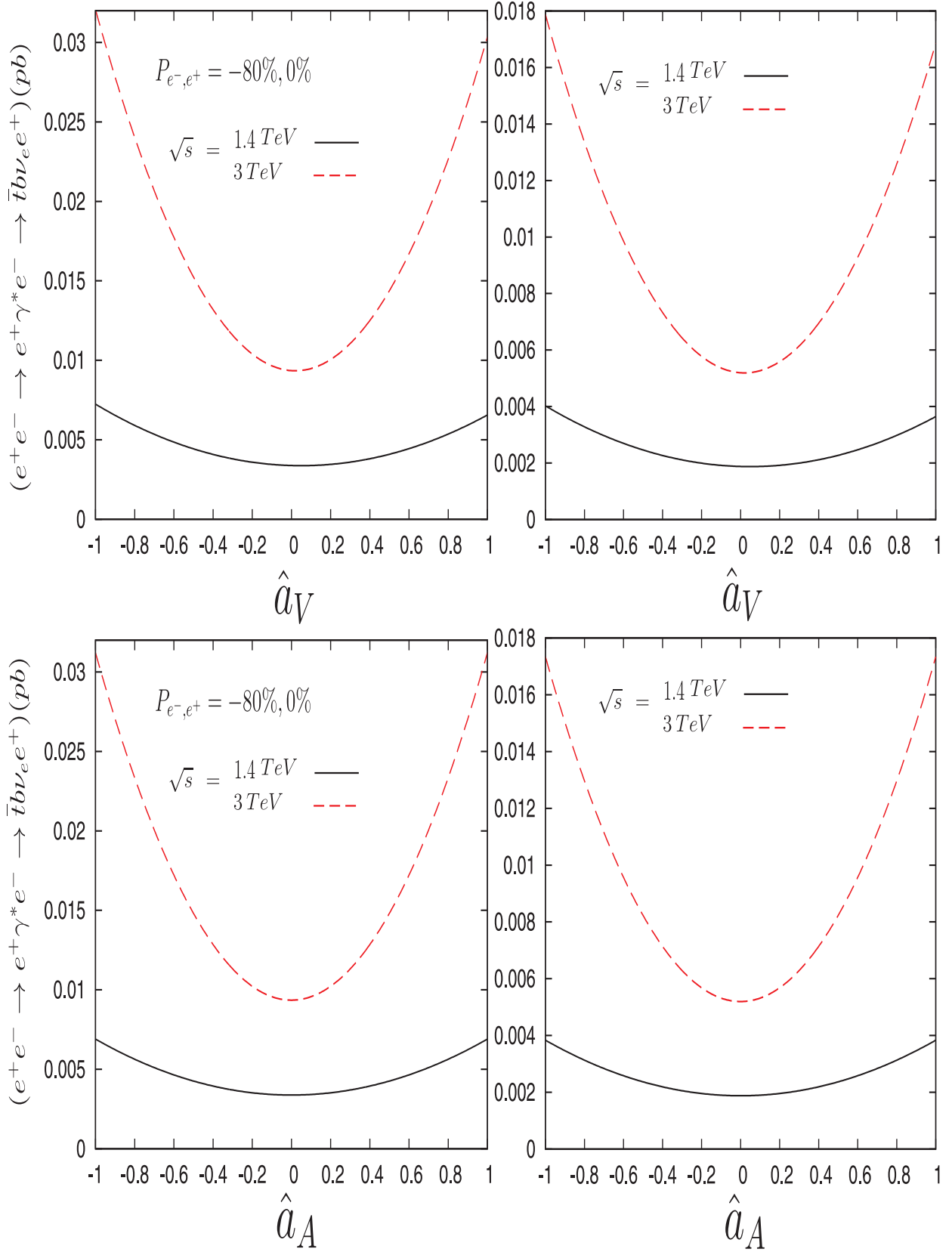}}}
\caption{\label{fig:gamma16} The integrated total cross section of the process
$e^+e^- \to e^+\gamma^* e^- \to \bar t b \nu_e e^+$ ($\gamma^*$ is the Weizsacker-Williams photon)
as a function of $\hat a_V$ and $\hat a_A$ with $\sqrt{s}=1.4, 3\hspace{0.8mm}TeV$, $Q^2=2$\hspace{0.8mm}$GeV^2$
and polarized and unpolarized beams.}
\end{figure}

\begin{figure}[t]
\centerline{\scalebox{0.8}{\includegraphics{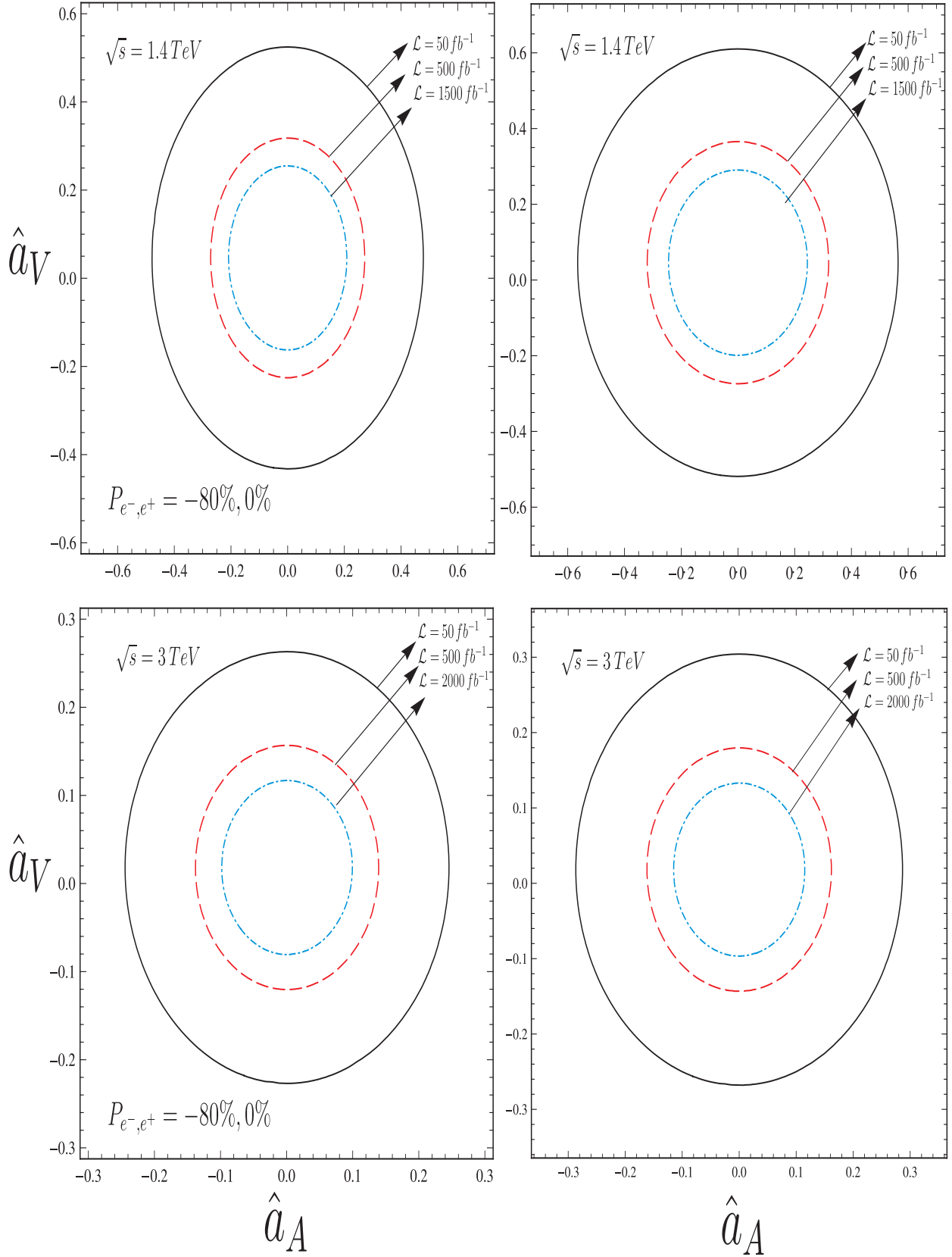}}}
\caption{\label{fig:gamma17} Limits contours at the $95\% \hspace{1mm}C.L.$ in the
$\hat a_V-\hat a_A$ plane for $e^+e^- \to e^+\gamma^* e^- \to \bar t b \nu_e e^+$ ($\gamma^*$ is the Weizsacker-Williams photon)
with $\sqrt{s}=1.4, 3\hspace{0.8mm}TeV$, $Q^2=2$\hspace{0.8mm}$GeV^2$ and polarized and unpolarized beams.}
\end{figure}

\end{document}